# The Closer the Better: Similarity of Publication Pairs at Different Co-Citation Levels[*]

Giovanni Colavizza[1], Kevin W. Boyack[2], Nees Jan van Eck[3] and Ludo Waltman[3]

[1] *giovanni.colavizza@epfl.ch*
Digital Humanities Laboratory, École Polytechnique Fédérale de Lausanne (CH)

[2] *kboyack@mapofscience.com*
SciTech Strategies, Inc. (USA)

[3] *{ecknjpvan, waltmanlr}@cwts.leidenuniv.nl*
Centre for Science and Technology Studies, Leiden University (NL)

**Abstract**

We investigate the similarities of pairs of articles which are co-cited at the different co-citation levels of the journal, article, section, paragraph, sentence and bracket. Our results indicate that textual similarity, intellectual overlap (shared references), author overlap (shared authors), proximity in publication time all rise monotonically as the co-citation level gets lower (from journal to bracket). While the main gain in similarity happens when moving from journal to article co-citation, all level changes entail an increase in similarity, especially section to paragraph and paragraph to sentence/bracket levels. We compare results from four journals over the years 2010-2015: *Cell*, the *European Journal of Operational Research*, *Physics Letters B* and *Research Policy*, with consistent general outcomes and some interesting differences. Our findings motivate the use of granular co-citation information as defined by meaningful units of text, with implications for, among others, the elaboration of maps of science and the retrieval of scholarly literature.

**Introduction**

The co-citation relation is used extensively in bibliometrics, and has received some recent attention in information retrieval. Applications include the identification of topically-related publications for search engines and clustering of publications to understand the structure of science. If two or more publications are co-cited by a third one, they are generally assumed to be related to some extent, from the viewpoint of their citing authors (Small 1973). Normally, this assumption is considered to be valid already at a relatively coarse co-citation level, most often at the publication (e.g. article) level. In addition, recent work suggests that the relatedness of co-cited publications might increase with increasing proximity of two publications within the full text of the citing publication (e.g. Gipp & Beel 2009), and that improvements in maps of science or document retrieval can be obtained by taking textual proximity into account (e.g. Boyack et al. 2013). Indeed, it makes sense to assume that if two publications are co-cited within the same sentence or bracket in a publication, they typically will be in some way more related than two publications co-cited only at the more general section or publication levels. Yet open questions remain. We know little about the ways in which related, co-cited publications are similar over different dimensions. Furthermore, to what extent do different notions of similarity, such as textual and intellectual, depend on the level of the co-citation? This study was designed to provide answers to these questions.

---

The recent increase in availability of full text data has the potential to impact research in bibliometrics and in the retrieval of scientific publications. In this study we explore the implications of co-citations at different levels of granularity, and we do so by: a) considering different measures of similarity: textual similarity, reference overlap, author overlap, and proximity in publication time; b) comparing articles from four journals representing different fields and possibly different citation practices and behaviors. The four journals considered in this study are *Cell*, the *European Journal of Operational Research*, *Physics Letters B* and *Research Policy*. After discussing related literature in the following section, we define our dataset and methods, discuss empirical results, and then conclude with implications and suggestions for future research.

**Related Work**

*Co-citation analysis*
Co-citation relations have been used in a variety of contexts. Originally introduced in 1973 independently by Small (1973) and Marshakova Skaikevich (1973) as a complement to bibliographic coupling, co-citation analysis has most often been used to study the structure of science from the perspective of cited publications. A pair of publications is considered to have been co-cited if they appear together in the reference lists of one or more citing publications. The more often a pair of publications has been co-cited, the more related they are assumed to be. The notion of co-citation has been extended to cited authors (White & Griffith 1981) and journals as well (McCain 1991; Ding, Chowdhury & Foo 2000). While it was originally used only in relatively small datasets, in recent years it has been used to cluster millions of publications (Boyack & Klavans 2014; Klavans & Boyack 2017).

Co-citation has so far found limited use in the context of information retrieval and recommender systems for scientific literature. According to Beel et al. (2016), only 10% of the surveyed research-paper recommender systems use a co-occurrence method, and 16% a graph based method, typically relying on citation networks. Much fewer use citation relations. Examples of graph based methods using generic citation relations include Baez et al. (2011), He et al. (2010), Liang et al. (2011), Woodruff et al. (2000) and Zhou et al. (2008). More recently, a co-citation based metric was successfully applied to the challenge of ranking a large number of papers (120M) with respect to their query independent importance (Ribas et al. 2016). A few contributions that specifically use co-citations and full-text relatedness in the context of recommendation systems (Gipp and Beel 2009, Eto 2016, Schwarzer et al. 2016) will be discussed below. The main feat of co-citations for recommendation and retrieval is their focus on the broader relatedness of publications, not just their feature similarity. Yet, the fact that most publications are never or almost never co-cited is a major limitation in this respect.

*Full-text analysis*
In recent times, the increasing availability of the full text of scientific publications has enabled a rising interest in its use for the purposes of, among others, information retrieval, automatic summarization of publications and detection of citation polarity or motivation (Ding et al. 2014; Jha et al. 2016; Hernández-Alvarez & Gomez 2016). An example application is the automatic summarization of a set of research articles: Quazvinian and Radev (2008) and Quazvinian et al. (2013) proposed using the text of citing sentences to produce summaries of individual articles. More recently, it was also shown using a set of nearly four million publications from Elsevier journals that full-text features can sensibly enhance the prediction of the future impact of a publication, and more broadly that "it is well worth the effort to obtain the full text of scientific articles and to exploit the power of natural language analysis" (McKeown et al. 2016, p. 2686).

Recent work has also been done using sentences from full-text publications. For example, pairs of articles that are cited within the same sentence have been shown to be more similar than pairs of articles that are cited within the same article (Tran et al. 2009), which has obvious applications in information retrieval and document clustering. Liu et al. (2014), using several hundred thousand articles from PubMed Central, found that queries against citing sentences do very well at finding highly relevant older articles, but not for newer ones. Doslu and Bingol (2016) used citation contexts (text surrounding the reference marker) to improve the ranking of items in a retrieval context using data from CiteSeerX. Citation contexts have also been used for sentiment analysis (Small 2011) and for identifying biomedical discoveries (Small, Tseng & Patek 2017). The use of full-text is also a bedrock in supporting the retrieval of scholarly literature. A recent survey of the literature on research-paper recommender systems found that 55% of the methods in the literature rely on some form of content-based filtering, using mostly word-based features (Beel et al. 2016).

*Relatedness based on co-citation and full text*
More closely related to our study are a number of works aimed at improving measures of publication relatedness using full text. The majority of these have investigated co-citation proximity, while a couple are based on bibliographic coupling.

Nanba et al. (2000) were the first to test if co-citation proximity was related to increased textual similarity. Using a small set of citation areas (typically a couple of sentences) extracted by hand, they reported a rise in the textual similarity of two articles as their co-citation proximity increased. Elkiss et al. (2008) explored a set of nearly 2500 full-text articles from PubMed Central, finding that the cosine similarity of articles increased if they were co-cited at closer granularities, considering both their abstracts and body texts. Gipp and Beel (2009) suggested weighting the links between pairs of articles in a co-citation network according to their co-citation level, using full weight for the sentence level, half weight for the paragraph level, etc. They found that this weighting scheme performed twice as well at retrieving relevant publications than an approach that does not account for link weights. As mentioned above, Tran et al. (2009) found that pairs of articles that are co-cited within the same sentence were more similar than pairs of articles that are co-cited only at the article level. Callahan et al. (2010) suggested using the structure of the citing publication to establish co-citation relations between publications, albeit without assessing their method with more than anecdotal evidence. Liu and Chen (2012) examined a large dataset of publications and resulting co-citation pairs from BioMed Central (BMC), analyzing the sentence, paragraph, section and article levels. They found that in general lower co-citation levels are correlated with higher co-citation frequency, supporting the use of co-citation level information if available, and suggesting that co-citations at the sentence level form the basic structure of the general co-citation network. Boyack et al. (2013) used co-citation proximity information, as mapped by character offset, in order to improve co-citation clustering of articles. Their results on nearly 300,000 full-text articles published in Elsevier journals in 2007 show that the textual coherence of resulting clusters can increase by up to 30%, at the price of a decrease in the size of clusters. Eto (2016) proposed instead a method to expand co-citation networks in support for literature searches, by establishing "rough" co-citation relations if two articles are cited in similar citation contexts. Finally, in a work using Wikipedia entries rather than scientific articles, Schwarzer et al. (2016) found that citation proximity analysis improved recommendation quality as compared to a simple co-citation approach.

Despite the above efforts, and the evidence in support of the idea that lower co-citation levels entail an increased similarity between publications, not much is known on what kind of similarity is involved, nor if the effect is similar in different fields of research.

**Methods and Data**

*Co-citation levels*

Our co-citation hierarchy starts at the journal level. The levels at which a pair of publications can be co-cited are, from high to low:
1. Journal: publications co-cited within the same journal. In our analysis, we consider articles as co-cited within the same journal if they are cited in the same journal in the same year.
2. Article: publications co-cited within the same article.
3. Section: publications co-cited within the same section – a logical unit of the publication identified by some header – in an article.
4. Paragraph: publications co-cited within the same paragraph in an article. Paragraphs are usually identified with some layout expedient such as indentation or interlinear space.
5. Sentence: publications co-cited within the same sentence in an article.
6. Bracket: publications co-cited at the same location in an article. These are often delineated with brackets or parentheses.

Note that a pair of publications co-cited at a given level is also considered as co-cited at any higher level. For instance, if two publications are co-cited at the paragraph level, then they are also considered as co-cited at the section, article and journal levels. In the literature co-citation proximity often, but not always, entails considering the textual distance between co-cited publications, such as character offset (e.g. Boyack et al. 2013). In our case we do not take this approach. We use instead a set of well-defined textual units (from article to bracket) to identify different co-citation levels. The two approaches are related but not identical. We therefore refer to co-citation levels instead of co-citation proximity. We order co-citation levels from high (journal) to low (bracket). In this study we consider only journal articles indexed in Web of Science as co-cited publications for analysis.

*Data*

Our analysis is based on a database that we have constructed of the full text of articles in journals published by Elsevier. In particular, the analysis considers the following four Elsevier journals: *Cell*, the *European Journal of Operational Research* (*EJOR*), *Physics Letters B* (*PLB*) and *Research Policy (RP)*. The analysis therefore includes journals from a variety of disciplines, possibly with quite different citation practices. *Cell* is a prominent life sciences journal. *EJOR* can be seen as a journal at the interface between the social sciences, computer science, and mathematics. *PLB* is a physical sciences journal. *RP* is a predominantly quantitative social science journal. The four journals have also been chosen because they are all relatively large, because they all have at least a moderately high citation impact in their respective fields, and in some cases also because we are somewhat familiar with the journals. We analyze co-citations in the four above-mentioned journals in the six-year time period 2010-2015. We have chosen to consider a six-year time period to make sure that we have a sufficiently large number of data points per journal and to guarantee the reliability of our analysis.

Full-text data from the four journals mentioned above was collected by first using the Crossref REST API to identify all publications in the four journals in the time period 2010-2015. Next, the Elsevier ScienceDirect API (Article Retrieval API) was used to download the full text of the publications in XML format.

In the parsing of the XML formatted full text of a publication, an important issue is the identification of sections, paragraphs, and sentences. For each in-text citation, we aimed to identify the section, paragraph, and sentence in which the reference is located. Sections, subsections and paragraphs could be identified directly using XML tags as encoded by Elsevier. In the case of sections, only main sections were taken into account. Subsections were ignored. Sentences could not be identified using XML tags. To identify sentences, we used a sentence splitting algorithm. The algorithm that we used is a modified version of the algorithm provided in the BreakIterator class in the Java 8 API. Sometimes multiple in-text citations are located within the same brackets in the full text of a publication. This was also detected by the algorithms that we used for parsing the full text of a publication. This means that for each in-text citation in a publication our algorithms were able to identify the section, paragraph, sentence, and bracket in which the reference is located. In addition, our algorithms also identified in-text citations occurring at various special locations in the full text of a publication, such as in footnotes and in the captions of tables and figures. Such citations have not been used in this study.

The dataset of pairs of co-cited articles was constructed as follows. First, we used every article published from 2010 to 2015 in the journals under consideration classified as either "full research article" or "short communication". Review articles were excluded since their specific co-citation patterns may be different from articles describing original research. Next, we restricted the cited articles to those which are indexed in Web of Science (WoS), without further filtering. Matching of references in citing articles with cited articles indexed in WoS was performed based on the combination of the name of the first author, the publication year, the volume number, and the first page number. A match was required for all four fields. In the case of matching based on the name of the first author, only the last name and the first initial of the author were taken into account. We made use of the in-house version of the WoS database of the Centre for Science and Technology Studies at Leiden University. We note that this database does not include publications from before 1980.

We defined six sets of co-cited article pairs for every journal, corresponding to the six co-citation levels defined above. At the article level and the lower levels, all co-cited article pairs were used. Co-cited article pairs at the journal level were sampled from the large number of possible pairs. For each pair of articles co-cited within the same article, the most detailed level at which it is co-cited was identified, and the pair was counted for that level and all higher levels up to the article level. The frequency of co-citations was not considered. Consider a pair of articles that have been co-cited multiple times at the article level (because at least one of the co-cited articles has more than one in-text citation). Suppose that one of the co-citations is at the bracket level. The pair of articles will then be considered as co-cited at the bracket level and consequently also at all higher levels.

The hierarchy of the number of pairs per level is shown in Table 1; for instance, *Cell* has 105,984 pairs at the sentence level of which 59,160 are also at the bracket level. For the journal level, we created a large set (2M) of unique pairs of co-cited articles for each journal by sampling with replacement from the available pool of articles that have been cited by articles from the given journal in the same year. The number of pairs maintained at the journal level is comparable to the numbers at other levels. Characterization of the dataset at all levels for each journal is given in Table 1. It must be stressed that our journals have a different coverage in WoS, as shown by the proportion of references indexed in WoS. Nevertheless, more than half of the references are indexed in WoS for every journal. We note in passing that both the average number of references per article and the Price Index vary significantly among journals, most notable in articles from *RP* having longer reference lists and older cited literature.

**Table 1: Basic statistics of the dataset for each journal. * indicates sampled data. ^ if calculated over citing articles in WoS. The Price Index is the proportion of references to articles published at most 5 years before the publication of the citing article (De Solla Price 1970).**

|  | Cell | EJOR | PLB | RP |
|---|---|---|---|---|
| # Citing articles (% in WoS) | 2,038 (98.6) | 3,550 (97.2) | 5,070 (97.8) | 767 (98.2) |
| # References (% in WoS) | 98,212 (95.9) | 122,137 (63.3) | 192,900 (71.3) | 54,322 (56.5) |
| References per article (mean/median)^ | 48.9 / 48 | 35.4 / 31 | 38.9 / 35 | 72.1 / 68 |
| # In-text citations (% in WoS) | 177,856 (93.2) | 176,703 (64.4) | 259,796 (72.2) | 80,138 (57.6) |
| In-text citations per article (mean/median)^ | 87.3 / 84 | 49.8 / 43 | 51.3 / 46 | 104.4 / 94 |
| Price Index^ | 0.52 | 0.35 | 0.43 | 0.27 |
| # Co-cited pairs – Journal* | 498,566 | 289,609 | 673,438 | 330,856 |
| # Co-cited pairs – Article | 2,501,398 | 866,603 | 1,515,834 | 570,127 |
| # Co-cited pairs – Section | 1,210,682 | 518,031 | 1,013,966 | 282,232 |
| # Co-cited pairs – Paragraph | 362,295 | 168,632 | 415,076 | 59,077 |
| # Co-cited pairs – Sentence | 105,984 | 76,834 | 210,694 | 23,079 |
| # Co-cited pairs – Bracket | 59,160 | 46,816 | 146,789 | 14,950 |

*Similarity measures*

The similarity between two publications can be conceptualized in many different ways. Here we focus on four main axes: textual similarity, intellectual overlap, author overlap and time distance. We have explored several other axes, such as publication venue (journal), the typology of the publications, and various counts (e.g. number of words, sentences and paragraphs), but most of them were relevant only for a very limited set of pairs, thus we decided to discard them from the analysis.

To establish the *textual similarity* between two publications we considered their titles and abstracts and used the BM25 measure, widely adopted to rank documents for information retrieval given a textual query (Spark Jones et al. 2000a, b) and more recently also used to cluster publications based on their titles and abstracts (Boyack et al. 2011). Each publication (title and abstract) was reduced to lowercase and split into tokens for calculation of textual similarity (eliminating punctuation and tokens of just one alphanumeric character). Given a publication $q$ and another publication $d$, BM25 similarity was calculated as

$$s(q,d) = \sum_{i=1}^{n} IDF_i \frac{n_i(k_1+1)}{n_i+k_1\left(1-b+b\frac{|D|}{|\overline{D}|}\right)}, \quad (1)$$

where $n$ denotes the number of unique tokens in $q$ and $n_i$ equals the frequency of token $i$ in $d$ ($n_i = 0$ for tokens that are in $q$ but not in $d$). The parameters $k_1$ and $b$ were set to the commonly used values of 2 and 0.75 respectively. $|D|$ denotes the length of publication $d$, in number of tokens. $|\overline{D}|$ denotes the average length of all publications in the collection. The IDF value for every unique token in the collection was calculated as

$$IDF_i = \log\left(\frac{N - d_i + 0.5}{d_i + 0.5}\right), \qquad (2)$$

where $N$ denotes the total number of publications in the collection and $d_i$ denotes the number of publications containing token $i$. IDF scores strictly below zero were discarded to filter out very commonly occurring tokens. BM25 is not a symmetric measure. We obtained a symmetric measure for the similarity of publications $q$ and $d$ as follows:

$$\frac{s(q,d) + s(d,q)}{2}. \qquad (3)$$

The BM25 textual similarity was calculated for every article pair listed in Table 1. For each journal, we then divided by the maximum BM25 value among pairs from that journal in order for the measure to range between 0 and 1. We note that after applying this normalization BM25 values cannot be directly compared between journals.

We define the *intellectual overlap* to be the proportion of references that a pair of publications have in common (alternatively known as bibliographic coupling):

$$\frac{N_{qd}}{\min(N_q, N_d)}, \qquad (4)$$

where $N_{qd}$ denotes the number of overlapping references of publications $q$ and $d$ and $N_q$ and $N_d$ denote the total number of references in publications $q$ and $d$ respectively. The intellectual overlap equals one if all references of the publication with the shorter reference list are also cited by the other publication. In the calculation of (4), all references were taken into account, not just references to articles indexed in WoS, and we required that two references are perfectly identical for a match to be established.

We define the *author overlap* as the proportion of authors that a pair of publications have in common. Analogous to the intellectual overlap measure we used the minimum number of authors of the two publications as the normalization basis (similar to Eq. (4)). If all authors of the publication with fewer authors are also authors of the other publication, the author overlap equals one. Matching of authors was done using the following process. Starting with a string of text for the surname of an author and one for the full name of the author, we converted both strings to lowercase. We then compared all possible combinations of author mentions. We considered two mentions to refer to the same author if the surname was identical and all components of the full name of the shorter name string had a match in the longer name string. For example, "John J. Abrams" will match with "J. Abrams" but not with "J. M. Abrams". Given that we operate within the context of a specific journal and that the procedure is fairly conservative, we deem the results as sufficiently accurate.

Lastly, we define the *time distance* as the number of months between the publication dates of a pair of co-cited publications. We used the publication month data present in WoS. Some articles do not have a publication month. For these articles only the year of publication or the season of publication (e.g., spring or summer) may be available. We then estimated the publication month. An article published in spring for instance is treated as an article published in March, and an article for which only the year is available is considered to have been published in June.

## Results

*Mean and median similarity*

We start our analysis by considering the mean and the median of the similarity measures of pairs of articles co-cited at different levels for each journal under consideration. Figure 11 shows the results for textual similarity calculated using the BM25 measure. Clearly, textual similarity monotonically increases as the co-citation level goes down. The means are located above the medians, suggesting that the distributions of the similarities are somewhat right skewed. A closer inspection of the distributions of the similarities (not shown here) reveals that the distributions are fairly normal, but indeed with a slight skew on the right side. For all four journals the largest increase in similarity is obtained by moving from the journal to the article co-citation level. The section level adds little afterwards. Behaviours then differ: whilst article pairs co-cited in *EJOR*, *PLB* and *RP* gradually increase in similarity, this increase is significantly more pronounced for *Cell*, especially when moving from the paragraph to the sentence level.

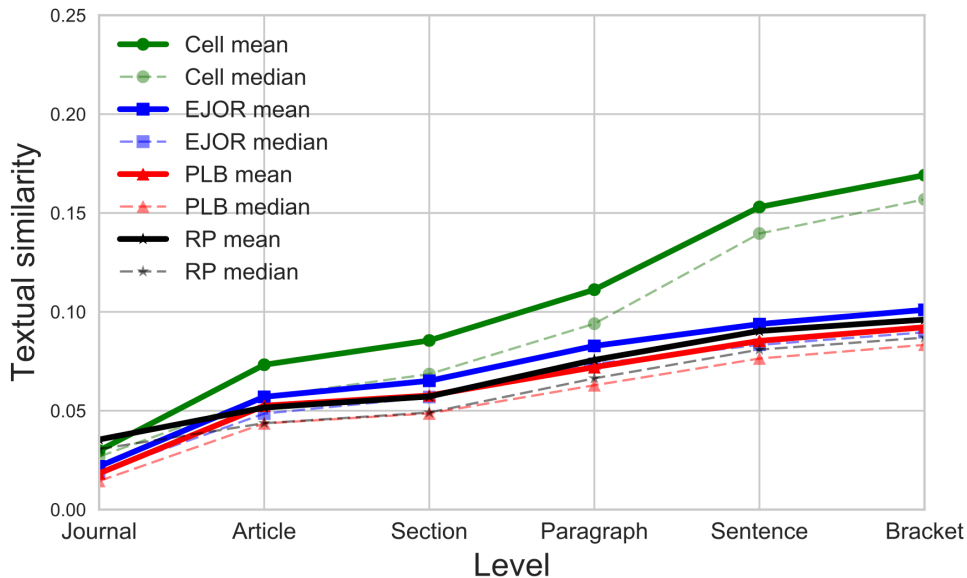

**Figure 1: Mean and median textual similarity (BM25) of article pairs co-cited at different levels.**

We consider next the intellectual overlap of pairs of articles with results shown in Figure 22. Intellectual overlap presents a generally more skewed distribution, and an even larger gain in similarity takes place from the journal level, where similarity is essentially zero, to the article level. Here too, the similarity rises monotonically as we move down in the co-citation levels. *PLB* and *EJOR* have the highest intellectual overlap in co-cited articles. The rise in intellectual overlap is also more rapid for *PLB*, ending up at the bracket level with an average number of nearly one in every four references being shared. Pairs of articles co-cited in *Cell* and *RP*, as well, present a rapidly growing overlap as the level lowers.

Results for author overlap are given in Figure 33. The distributions are very skewed, as at any level most pairs of co-cited articles do not share authors at all. Nevertheless, we see that author overlap in the case of *PLB* is stronger than in the case of the other journals. We also note, based on Table 2, that articles published in and cited by *PLB* have a more skewed distribution of the number of authors than those of *Cell*, *EJOR* and *RP*. Physics is, in general, known to have a larger number of kilo-authored articles than other fields. *Cell*, on the other hand, has a higher median number of authors per article, as shown in Table 22, which might lower its relative author overlap. It is worth noting that for all journals there is a rapid growth in author overlap when moving from the section to the paragraph level and from the paragraph to the sentence level.

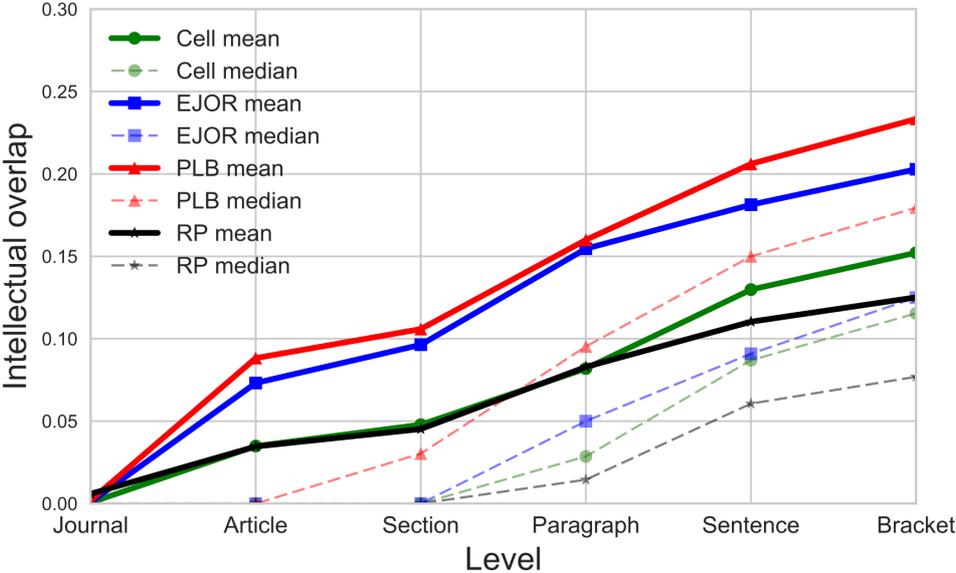

**Figure 2: Mean and median intellectual overlap of article pairs co-cited at different levels.**

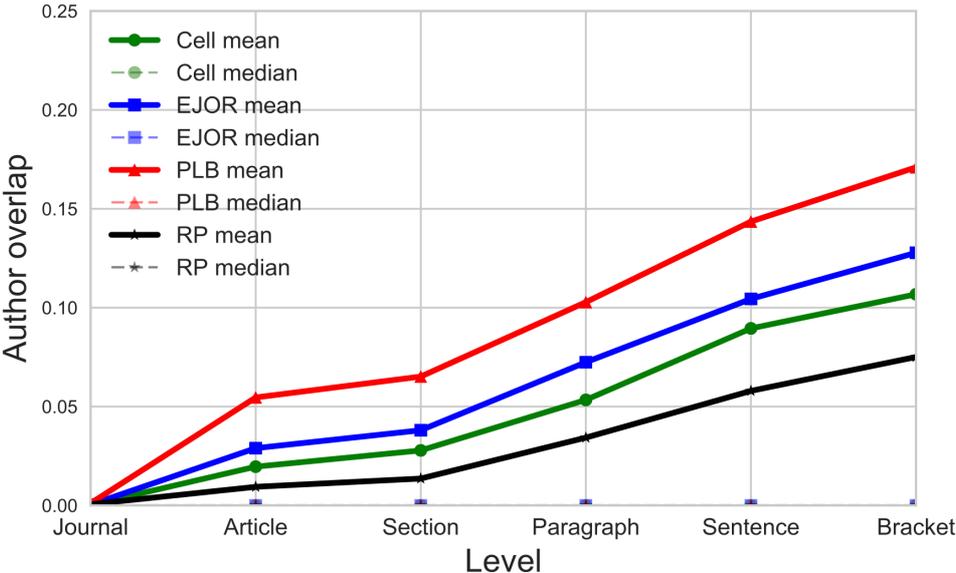

**Figure 3: Mean and median author overlap of article pairs co-cited at different levels.**

**Table 2: Mean (and median) number of authors of the citing and cited articles.**

|  | **Cell** | **EJOR** | **PLB** | **RP** |
|---|---|---|---|---|
| **Citing articles** | 11.6 (9) | 2.7 (3) | 133 (3) | 2.5 (2) |
| **Cited articles** | 6.4 (5) | 2.4 (2) | 38.4 (3) | 2.1 (2) |

Lastly, in Figure 44 we consider the distance of the publication time for pairs of co-cited articles. *Cell* in general co-cites articles more proximal in time as evidenced by its lower time distances at every level. As we move down in the co-citation levels, co-cited articles become increasingly similar in terms of their publication time, ending up at a median difference in publication time of about three years for *Cell* and *PLB* and four years for *EJOR* and *RP* at the bracket level.

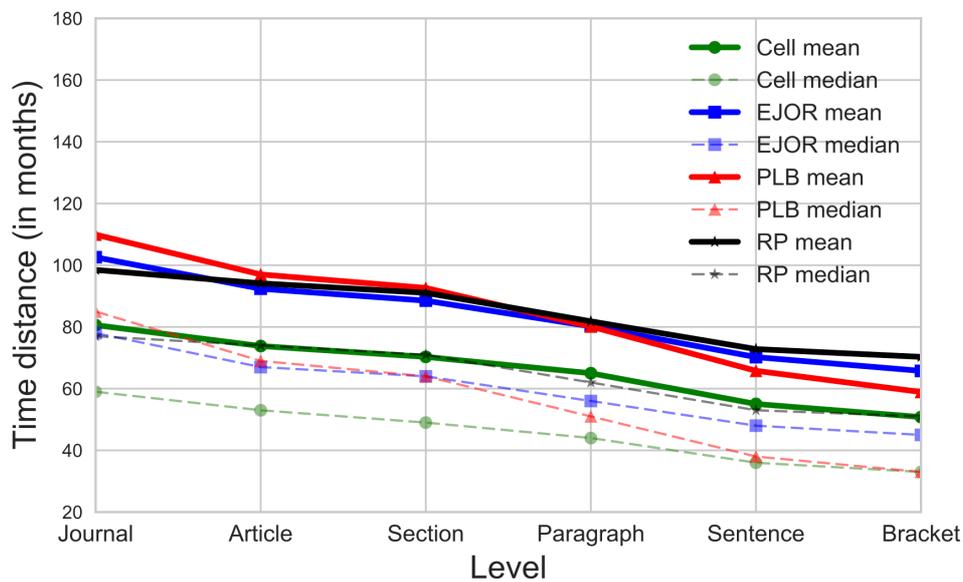

**Figure 4: Mean and median publication time distance (in months) of article pairs co-cited at different levels.**

Taken together these results highlight how over four different dimensions (text, references, authors and time) co-cited articles that are located at lower levels tend to be more similar than those that are located at higher levels. Results confirm the intuitive idea that, as the co-citation level lowers, research being co-cited is increasingly more related over a variety of dimensions. The agreement of all similarities supports the idea that, especially at lower co-citation levels, authors tend to refer to coherent pieces of research that are similar in content, sometimes also similar in terms of authors involved, and that were published around the same time. Some differences are worth pointing out. *Cell* pairs seem to become more textually similar and to published closer in time, whilst *PLB* and *EJOR* pairs are more intellectually similar and share a higher proportion of authors as the level lowers.

*Cumulative distributions*
We now take another view by considering cumulative distributions. For each journal, each similarity measure and each proximity level, we inspect the proportion of pairs of co-cited articles with a similarity value equal to or below a certain threshold value. All results are shown in Figure 5. Starting with textual similarity, we can immediately appreciate how gains in similarity are not negligible at most levels. This is especially the case for pairs of articles co-cited in *Cell*, which seem to increase their similarity in three steps: from journal level to article/section, to paragraph and to sentence/bracket. For *EJOR, PLB* and *RP* these steps are less well defined. The main increase takes place between the section and paragraph levels for these journals.

The cumulative distribution of the intellectual overlap follows slightly different patterns. The similarity of pairs of co-cited articles according to intellectual overlap is indeed more pronounced in *PLB* than in the other three journals. Three significant increases in similarity can be observed for *Cell, PLB* and *RP*, namely between the section and paragraph levels, between the paragraph and sentence levels, and between the sentence and bracket levels, whilst for *EJOR* the main significant increase takes place between the section and paragraph levels.

Lastly, we consider the same distributions for the author overlap and the publication time distance. With respect to the author overlap, we notice how pairs of co-cited articles which share authors are relatively rare at all levels, as was indeed evident already from Figure 33, and the importance of this aspect for *PLB* is particularly remarkable. In a similar way to what we discussed for the intellectual overlap, the author overlap seems to follow a three step pattern of increase for *Cell, PLB* and *EJOR*. The distributions for the publication time distance are more similar at different levels, except for *PLB,* where gains in time distance moving to lower levels are somewhat more important.

*Precision/recall*
In order to explore the discriminative power of the similarity measures under consideration, we analyze their capacity to distinguish between pairs of articles co-cited at the journal level and pairs of articles co-cited at other levels. We defined a basic classification task by sampling a set of pairs at the journal level, which we consider to be negative cases, and an equally sized set of pairs at another lower level of interest, which we consider to be positive cases. Our aim is to use a similarity measure to distinguish between the positive and the negative cases. We sorted pairs of co-cited articles according to a given similarity measure, and calculated the precision and recall as we move through the sorted list of pairs. Results for the four journals at the article, paragraph and bracket levels are given in Figure 6, by considering author overlap, intellectual overlap as well as textual similarity.

Precision and recall scores were calculated by considering a sample of 10,000 pairs at a given level and an equal-size sample at the journal level. Thus, each precision-recall analysis involved 20,000 pairs, of which half were positives and half were negatives. Pairs were sorted in decreasing order according to a certain similarity measure, with a random ordering of ties, and precision and recall scores were calculated by considering the pairs at the journal level as negative cases, and the pairs at the other level as positive cases. Precision and recall can both range between 0 and 1.

As can be seen in Figure 6, author overlap is able to achieve a high level of precision, but only with a very limited recall, faring slightly better in *PLB* than in *Cell*, *EJOR* and *RP*. Intellectual overlap has a much better precision-recall trade-off than author overlap, especially pronounced for *PLB, Cell*, and *EJOR* in this order. Textual similarity has the most gradual trade-off between precision and recall. For most levels of recall it has the highest precision, in particular in *Cell* and *EJOR*. Hence, intellectual overlap and textual similarity both perform well in distinguishing pairs of articles co-cited at the journal level from pairs of article co-cited at other levels, with textual similarity clearly outperforming intellectual overlap for two of the four journals (*Cell* and *EJOR*) and intellectual overlap faring better for the other two (*PLB* and *RP*). For all similarity measures the discriminatory power markedly improves as we move to lower co-citation levels. These results suggest that with respect to the task of retrieving relevant publications, different similarity measures may be preferred for different scholarly disciplines.

In general, it seems that the similarity of pairs is much stronger for *Cell*, *PLB* and *EJOR* as we move to more proximate co-citation levels. *RP* presents also notably low results in the most basic article co-citation level. We might speculate that this could be a consequence of the shorter half-life of publications in *Cell*, *PLB* and *EJOR*, according to the Price Index, where *RP* would fit a more social science profile with lower overall similarity between co-cited articles at all levels (Table 1). *RP* in this respect is an example of the citation practices of the social sciences, where collaborations are also less frequent and done with fewer co-authors (cf. Table 2), and there is less overlap in citations due to the fragmentation of the field (Hicks 1999). Another element for consideration is the likely influence of the different writing styles of different disciplines, at lower levels (paragraph to brackets, see e.g. Hayland 1999).

**Conclusions**

In this paper we explored the similarity of article pairs which are co-cited at different levels. We considered four journals from different disciplines: *Cell*, the *European Journal of Operational Research* (*EJOR*), *Physical Letters B* (*PLB*) and *Research Policy (RP)*. Our results indicate that the similarity of pairs of articles increases monotonically with their co-citation level. In other words, the lower the level at which two articles are co-cited, the more similar the articles will be on average. We used different measures of similarity between pairs of co-cited articles: textual similarity, intellectual overlap, author overlap and proximity in publication time. These measures all increase as the co-citation level lowers, with textual similarity and intellectual overlap being the most discriminative of article pairs co-cited at the article or lower levels. A similar study, based on a reduced dataset of co-citation pairs of articles from Elsevier, can be compared to the present one (Colavizza et al. 2017).

The four journals displayed some differences: *Cell* pairs show a higher relative gain in textual similarity as the co-citation level lowers, while *PLB* and *EJOR* pairs present a higher relative gain in intellectual and author overlaps. In absolute terms, *PLB* and *EJOR* article pairs are more similar in terms of intellectual and author overlap, *Cell* articles in terms of time distance. *RP* shows instead average or low absolute similarity across the same three measures. Furthermore, in most cases there are two transitions which entail a substantial relative increase in article pair similarity: article/section to paragraph to sentence/bracket, meaning that co-cited articles at these levels are sensibly more similar than articles co-cited at higher levels. This observation possibly relates to the way the state of the art is discussed. Under this interpretation, previous art discussed in textual units of increased narrowness is more related. Yet, not all textual units here considered are distinctive with respect to the discussion of previous art: the article/section and sentence/bracket levels respectively seem to discuss previous art at a comparable level of granularity.

Our results provide a solid confirmation of the idea that there is value in using co-citation proximity information in order to capture stronger links between co-cited articles. All co-citation levels provide meaningful information regarding the similarity of co-cited articles, but especially co-citations at low levels indicate a strong similarity of the co-cited articles. Our results support the use of co-citation level information, where available, to improve the performance of information retrieval systems and the accuracy of bibliometric analyses. Having shown that co-cited articles at lower levels are more related according to specific similarity measures, this insight can for example be used to improve the performance of literature retrieval or recommendation systems.

This study is not without limitations. In particular, it explores only four journals and thus cannot truly represent co-citation behaviours by discipline or field. Additional studies with larger datasets are needed to explore field-based differences. In addition, this study explores similarity measures independently without considering their mutual interactions. In other words, the effect of a measure on the relative change of other measures is not considered.

There are various next steps that could be taken. In addition to distinguishing between co-citations at different levels, full-text data also makes it possible to assign weights to direct citation links between articles. The weight of a citation link can be defined by the corresponding number of in-text citations. Weighted citation links could be studied in a similar way as we have done for co-citations at different levels. Another possibility could be to further explore how the results obtained for the different journals can be explained by disciplinary differences, both epistemic differences and differences in publication and citation practices, as well as to attempt to develop information retrieval systems and produce maps of science on the basis of fine-grained co-citation information.

**Acknowledgments**
Colavizza benefits from a Swiss National Fund grant, number P1ELP2_168489.

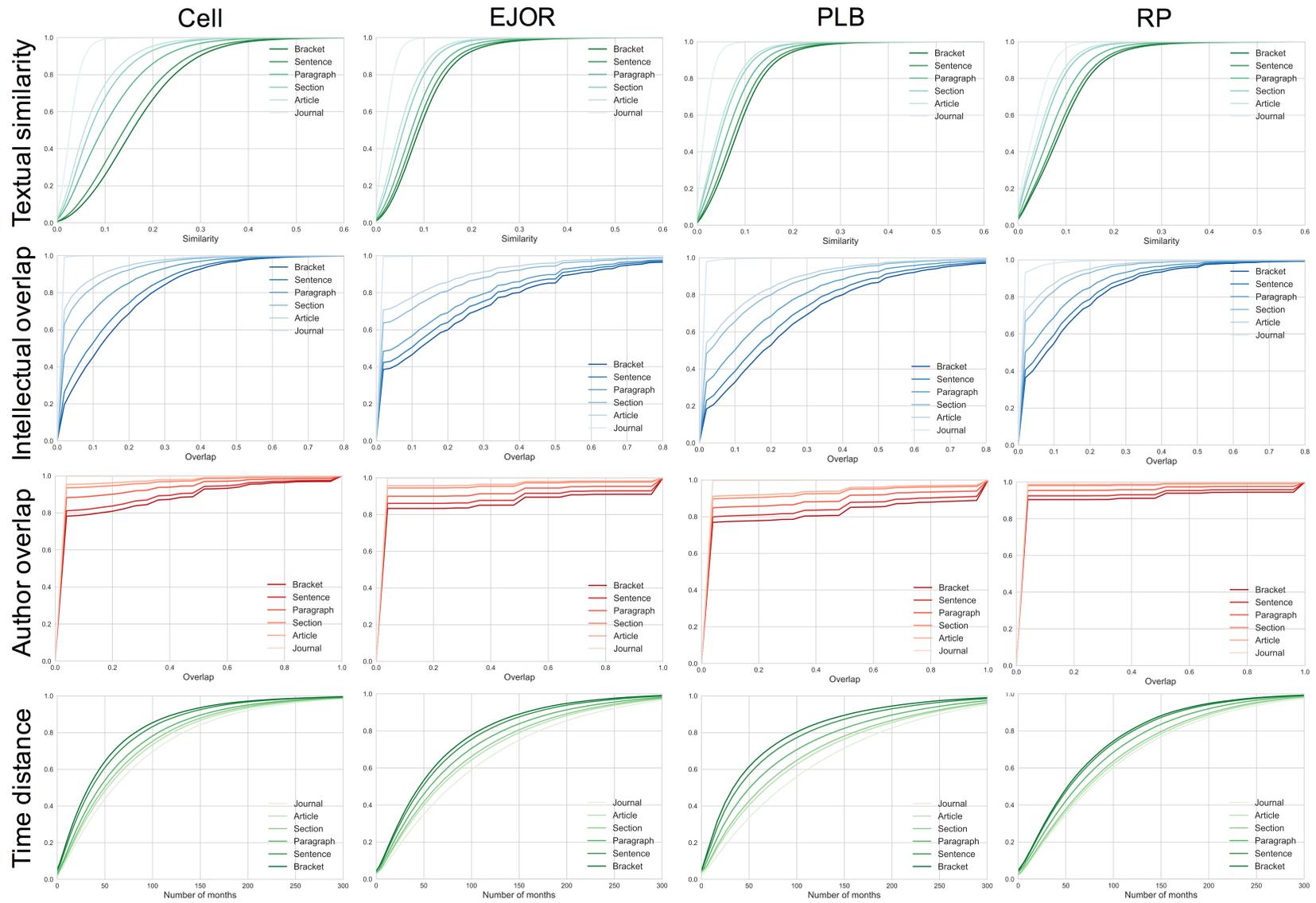

Figure 5: Cumulative distributions (y axis) of similarities of articles co-cited at different levels.

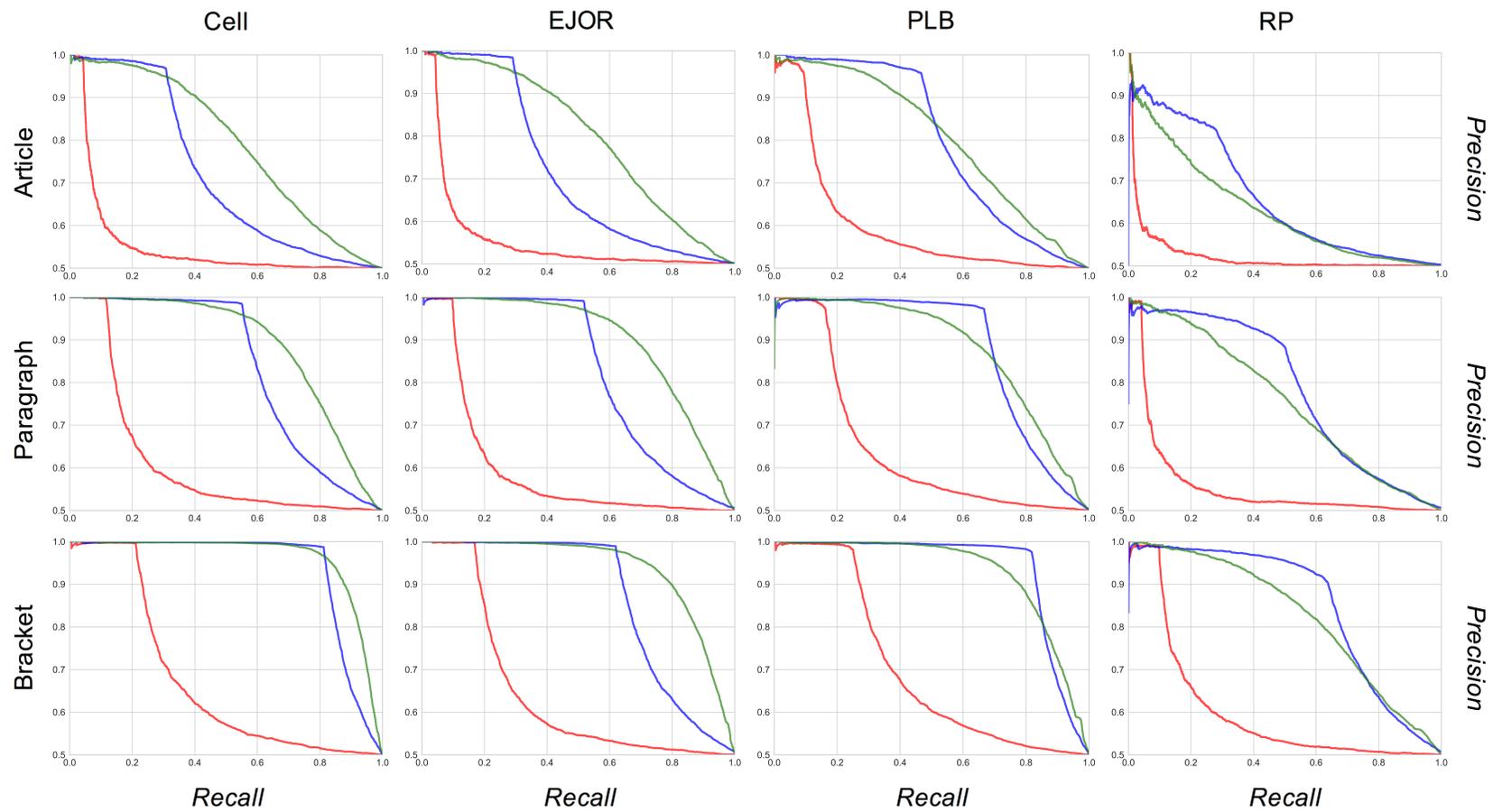

Figure 6: Precision (y axis) and recall (x axis) curves for each journal at the article, paragraph and bracket levels, for author overlap (red/grey), intellectual overlap (blue/dark grey) and textual similarity (green/light grey).

17